\documentclass[]{spie}  

\usepackage{amsmath,amsfonts,amssymb}
\usepackage{graphicx}
\usepackage[colorlinks=true, allcolors=blue]{hyperref}

\usepackage{xspace}
\newcommand{\Finesse}{\normalfont\textsc{Finesse}\xspace}

\title{A matter of perspective: how nanoscale optical defects limit cosmic-scale gravitational wave observations}

\author[a,b]{Anna C. Green}
\author[a,c]{Antonella Bianchi}
\author[d]{Daniel D. Brown}
\author[a,e]{Félice Feldmann}
\author[a]{Jeremie Gobeil}
\author[a]{Miron van der Kolk}
\author[a,c]{Riccardo Maggiore}
\author[a,c]{Jonathan W. Perry}
\author[a,b]{Emma Prins}
\author[a]{Mischa Salle}
\author[a,c]{Alina Soflau}
\author[a]{Enzo Tapia}
\author[a,c]{Andreas Freise}

\affil[a]{Nikhef, Netherlands}
\affil[b]{Maastricht University, Netherlands}
\affil[c]{Vrije Universiteit Amsterdam, Netherlands}
\affil[d]{The University of Adelaide, Australia}
\affil[e]{TU Delft, Netherlands}

\authorinfo{Corresponding author: A.C.G.;  E-mail: agreen@nikhef.nl}
 
\begin{document} 
\maketitle

\begin{abstract}
Ground-based gravitational-wave (GW) detectors, such as LIGO, Virgo, and KAGRA, have revolutionised astronomy. 
Now, future detectors like the Einstein Telescope and Cosmic Explorer aim to achieve even greater sensitivity. 
Advanced optical simulations are crucial to overcoming the challenges faced by these complex interferometers. 
\Finesse, the leading interferometer simulation tool in the GW community, supports the design and commissioning of these detectors by modeling optical, quantum, and mechanical effects. 
A key focus is understanding optical defects that distort the shape of the laser light and limit detector performance.
This work explores how nanoscale defects affect GW observations and presents recent advancements in modeling their effects to guide the development of next-generation detector optics.
\end{abstract}

\keywords{Interferometry, Gravitational Waves, Metrology, Numerical Simulations}


\section{Introduction}

\subsection{Gravitational Wave Astronomy}

Since their first direct detection in 2015\cite{GW150914}, gravitational waves (GWs) have emerged as an exciting new signal with which to observe the universe, and understand the fundamental physics driving it. 
These waves result from the interaction of accelerated massive objects with space-time, causing distortions which ripple out from the source event. 
The weak nature of this interaction means that only the most extreme events, such as the inspiral and merger of compact binary black holes observed in the first detection, can be detected. 

Current observations are led by a network of ground-based, kilometre-scale interferometers: LIGO\cite{AdvancedLIGO}, Virgo\cite{AdvancedVirgo} and KAGRA\cite{KAGRA} (LVK). 
These measure the phase difference accumulated as light propagates through the distorted spacetime, detecting characteristic strains $\Delta L / L$ of up to order $10^{-24}$\,Hz$^{-1/2}$, with a sensitivity band of c. 20Hz to a few kHz. 
Such facilities are complimented by other detection techniques, such as the International Pulsar Timing Array\cite{IPTA} which probes nanohertz signals, and the upcoming LISA mission\cite{LISAredbook}, which will measure the millihertz band.

Nearly 100 detections have now been confirmed by the LVK collaborations\cite{GWTC-1,GWTC-2,GWTC-2.1,GWTC-3}, as the observatories continually improve technologies which increase their observing range, observing time, and corresponding detection rates.
Of particular note is GW170817\cite{GW170817}, the first detection of binary neutron stars, which represents the beginning of gravitational-wave multi-messenger astronomy: electromagnetic counterpart signals were also detected, representing a significant collaboration by much of the global astronomy community\cite{GW170817-MM}.
Such signals are of particular interest for probing the behaviour of matter and fundamental forces in extreme environments that cannot be replicated on Earth.
The LVK has also detected a wide array of binary black hole events, which can uniquely be detected using GWs as they are not expected to produce counterparts, and other binary mergers (e.g. NS-BH), allowing the beginnings of population studies that give insights into how the universe evolved.

The GW community now plans to construct new ground-based interferometric detectors that will exceed the anticipated limits of the LVK facilities.
The proposed Einstein Telescope (ET, Europe)\cite{ETCDR,ET2020DRU,ET2025DL} and Cosmic Explorer (CE, USA)\cite{CEhorizon} projects will each exploit new and existing technologies, aiming to gain an order of magnitude in sensitivity, detecting signals from the entire visible universe, and a broader bandwidth, increasing the variety of potential signal sources\cite{ETbluebook, ETCoBA}.  
Targeting operation by the mid-2030s, each project is currently in an intensive period of R+D.

\subsection{Optical Defects in Gravitational-Wave Detectors}
\label{sec:Defects}

\begin{figure} [tb]
   \begin{center}
   \begin{tabular}{cc} 
   \includegraphics[height=8cm]{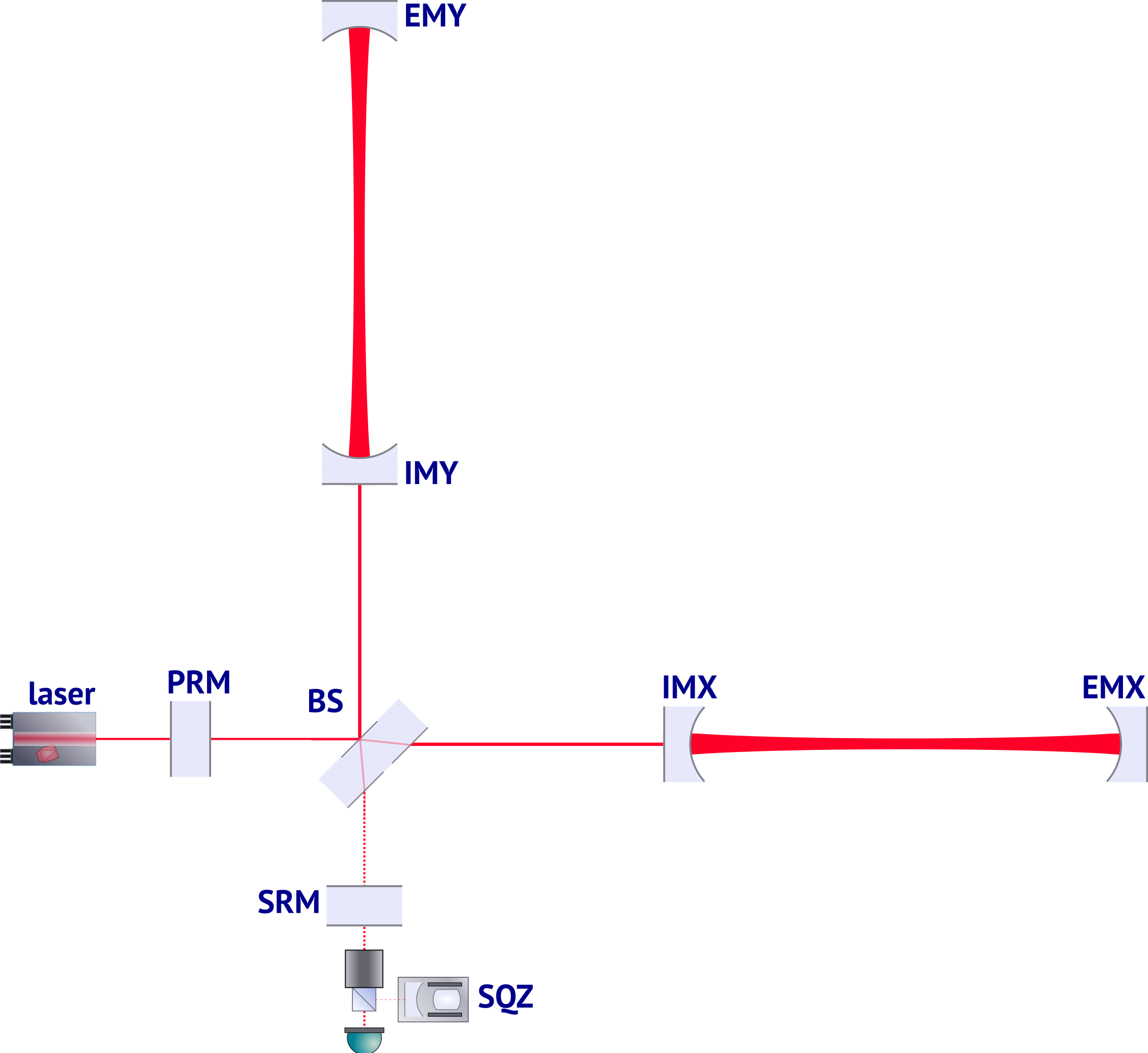} 
   \end{tabular}
   \end{center}
   \caption[GWD] 
   { \label{fig:GWD} 
   Schematic depicting the primary optical elements of modern interferometric ground-based gravitational wave detectors. 
   The configuration is referred to as a DRFPMi, a dual-recycled Michelson interferometer with Fabry-Perot arm cavities. 
    }
\end{figure}

Figure~\ref{fig:GWD} provides a conceptual overview of the key optical elements of all current and proposed ground-based interferometric GW detectors (GWDs). 
The configuration is referred to as a dual-recycled Michelson interferometer with Fabry-Perot arm cavities. 
the use of cavities inside each arm of the interferometer (X and Y in the figure) increases the interaction time with the GW signal. 
The mirrors forming these cavities (input and end mirrors, IMX/Y and EMX/Y) then serve as test masses, whose relative separation in spacetime is distorted by GWs.
The optic positions are carefully tuned to an operating point that maintains destructive interference at the output port; observing light thus indicates a GW signal.
Most light therefore returns towards the laser; the power recycling mirror (PRM) reflects this back into the interferometer, increasing the circulating power to reduce shot noise.
Finally, the signal recycling mirror (SRM, sometimes referred to as the signal extraction mirror) is used to shape the frequency response of the detector, and frequency-dependent squeezed light \cite{FDS-LIGO,FDS-Virgo} is injected to reduce quantum noise.

The parameters of the configuration determine the overall sensitivity and bandwidth of the detector, and corresponding astrophysical and cosmological observables:
Higher mass sources such as black holes tend to produce GWs at lower frequencies (Hz), while neutron star signals of interest are at higher frequencies (kHz). 
The final performance is limited by a wide array of noise sources - for example, seismic noise provides an ultimate lower frequency limit for ground-based detectors, therefore the optics are suspended. 
The major parameters of the optics themselves, particularly the IMs and EMs, play a significant role, leading to a challenging combination of requirements\cite{Degallaix2019}. 
The primary design aspects are driven by the performance in various frequency bands:
\begin{itemize}
    \item the overall frequency response of the optical system is determined by the reflectivities and curvatures of the optics and the resulting resonance condition.
    \item At high frequencies ($>100$\,Hz) the detector is \textbf{quantum shot noise} limited. 
    The noise scales inversely to the circulating power, $P$, thus high circulating power -- reaching hundreds of kilowatts in current LIGO detectors and targeting up to 3\,MW for ET's high-frequency-targeted interferometers \cite{ETCDR} -- and low optical loss is required. 
    The use of squeezed light to further mitigate quantum noise further increases the demand for very low loss optics.
    \item At mid-frequencies ($\sim100$\,Hz), \textbf{thermal (Brownian) noise} limits the performance. 
    This can be mitigated, to a degree, by using beam sizes \cite{Wang2018} and shapes\cite{Bond11,Tao2020} that distribute optical power over larger areas on larger optics.
    However, ultimately it is an intrinsic property of the optical material. 
    Thus there is an extensive global effort to develop new optical coatings such as crystalline AlGaAs \cite{Cole2023} or amorphous Ti-doped Germania \cite{Vajente2021} and substrate materials such as crystalline Silicon or sapphire that have low absorption, high Q-factor, low thermal expansion, and high thermal conductivity, as well as compatibility with cryogenics\cite{KAGRA,ETCDR}. 
    \item At low frequencies, \textbf{quantum radiation pressure noise} can limit the performance, scaling with $P/m^2$ where $m$ is the mass of the suspended optic. 
    Thus, current GWD optics weigh up to 40\,kg, and future detectors are considering masses up to 10 times larger\cite{CEhorizon}. 
\end{itemize}

Beyond these first considerations, special care must be taken to address optical defects, which can limit the detector performance.
The designs of current 
and proposed future GWDs
, rely on the use of split, coupled resonant optical cavities as depicted in Figure~\ref{fig:GWD}.
These are ideally designed for the fundamental Gaussian optical mode, however in practice the optics will include imperfections that distort the beam on reflection and/or transmission. 
The impact can broadly be categorised into two regimes (see Figure~\ref{fig:cav+dist}):

\begin{figure} [tb]
   \begin{center}
   \begin{tabular}{c} 
   \includegraphics[height=3.5cm]{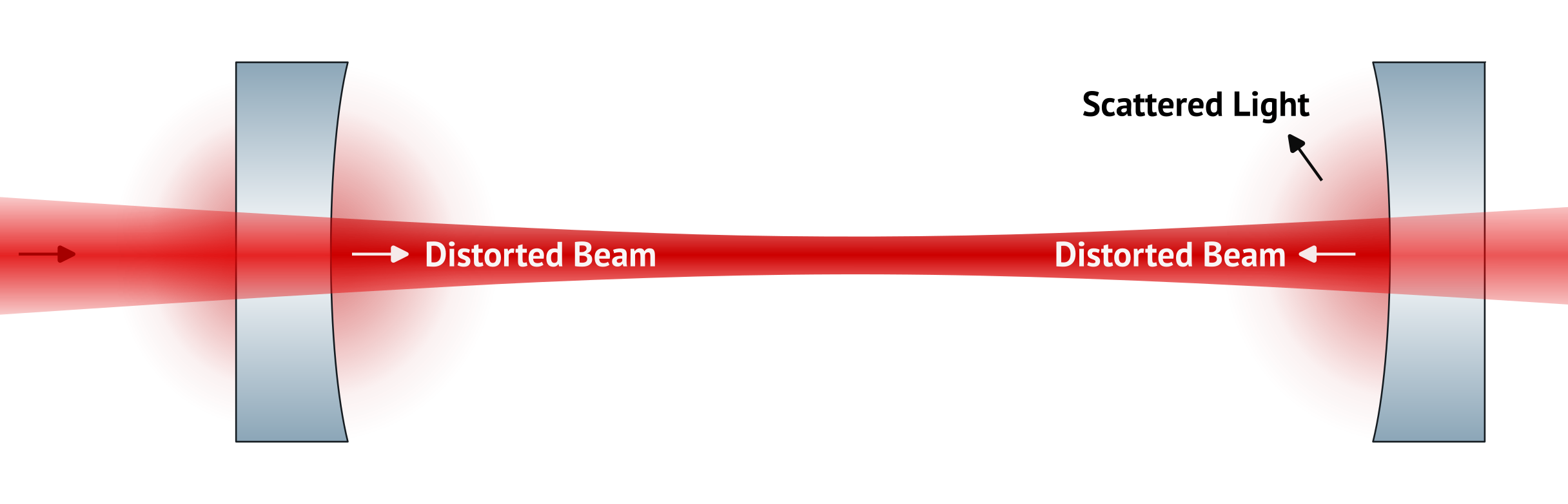} 
   \end{tabular}
   \end{center}
   \caption[Mirror in cavity] 
   { \label{fig:cav+dist} 
A mirror containing optical defects distorts the optical field upon reflection. 
When the optic forms part of a cavity, the resulting impacts can broadly be decomposed into wide-angle scattering which causes light to exit the optical system,
and small-angle distortions which remain in it. 
    }
\end{figure} 

\paragraph{Scattered Light} is reflected from an optic at wide angles and will \textit{exit the cavity}, no longer interacting with the optical system. 
The behaviour is typically due to surface micro-roughness.
This primarily contributes to the total optical loss, reducing circulating power and degrading squeezed light, so the GW signal amplitude is reduced. 
However, this light may reflect off other surfaces in the detector, such as the walls of the vacuum system or optical mounts, and find a path back into the main optical field. 
This can create significant challenges for signal analysis, as the light no longer takes the intended propagation path through the interferometer, so may have a different frequency dependence, and will inherit the environment noise of the surface(s) it interacts with.
This can create a signal in the GW detection band, masking genuine GW signals\cite{Tolley2023}.
Therefore, the surface microroughness should be minimized, with typical requirements for a surface RMS $<100$\,pm, and extensive steps are taken to baffle any light that does exit the interferometer\cite{Austin2020}.

\paragraph{Beam Distortions} -- the focus of this work -- instead considers imperfections introduced to the optical field that \textit{remain in the optical system}.
This includes defects that are intrinsic to individual optics, such as surface flatness resulting from polishing and coating processes, or substrate birefringence\cite{Wang24}, as well as defects that accumulate through the optical system, for example resulting in a frequency-dependent source of loss attributed to mode mismatch at the LIGO detectors\cite{McCuller2021,Toyra17}. 
The performance of current GWDs, particularly at low frequencies, is significantly limited by controls noise -- spurious noise from the environment that masks the GW signal by coupling into the interferometer via the control systems intended to keep it at its operating point. 
This is often associated with optical defects such as misalignment\cite{Maggiore25}.
Beam distortions may also be caused by defects that  depend on the incident optical power, resulting in effects such as thermal lensing  and point absorbers\cite{Brooks2021} which require extensive active mitigation\cite{Jones24}, or transient behaviours that overwhelm the control scheme such as parametric instabilities\cite{Evans15}.
\\
While the primary impact of such defects is often optical loss, since the distorted beam shape typically does not propagate in the optical system in the same way as ideal Gaussian, the referenced examples above highlight that the cause and impact of such optical defects can vary widely.
This typically requires detailed knowledge of the full interferometer to interpret and develop effective mitigation strategies.

\section{Simulating Optical Defects in GWDs}

\begin{figure} [tb]
   \begin{center}
   \begin{tabular}{c} 
   \includegraphics[height=7cm]{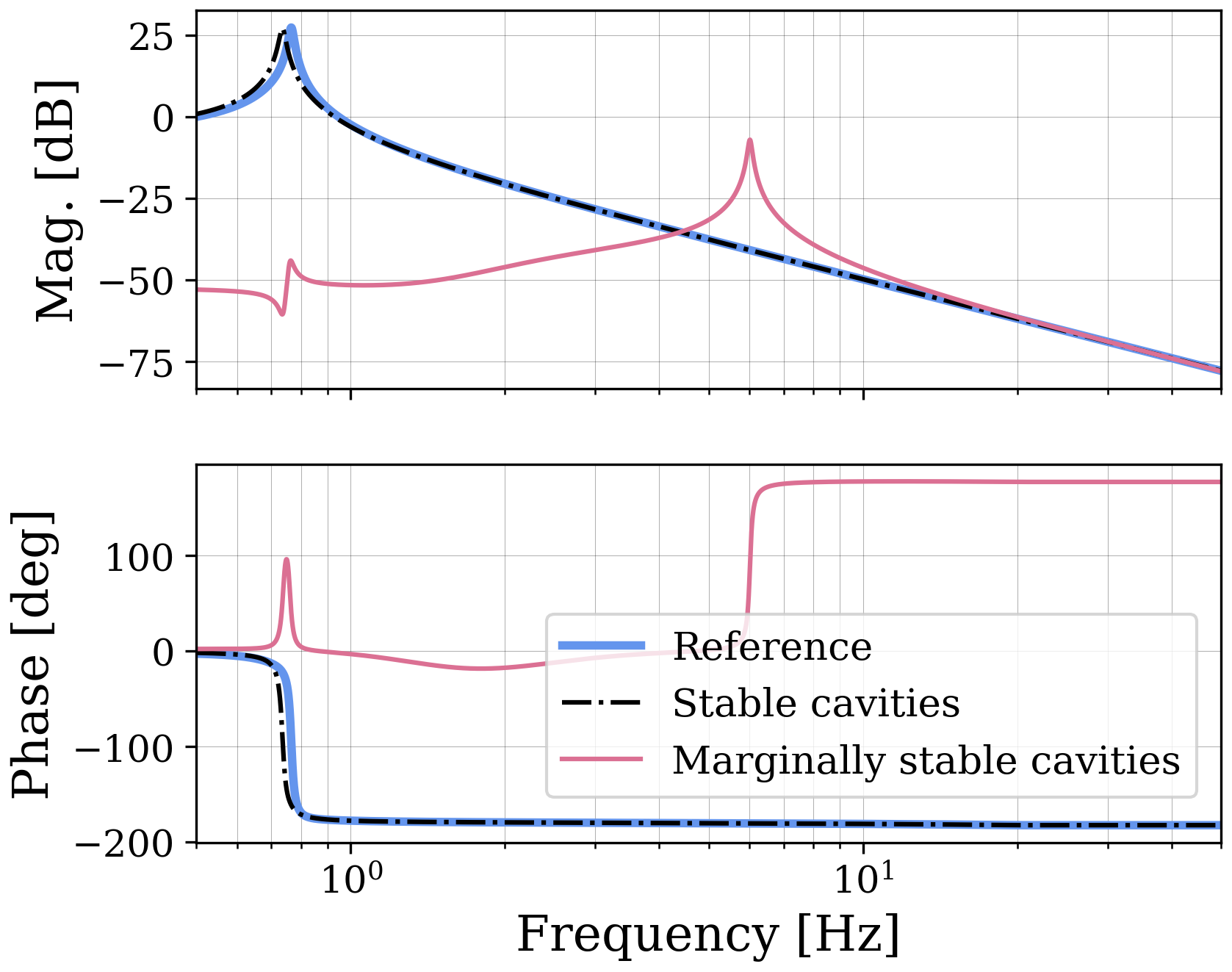}
   \end{tabular}
   \end{center}
   \caption[Offsets] 
   { \label{fig:offsets} 
   The transfer function of differential arm motion to the detection port of a gravitational wave detector is degraded by offsets in the control loops used to set the operating point of the detector. 
   These offsets result from small optical defects that alter the phase accumulated by the light fields composing the error signal as they propagate through the optical system. \\
   The optical configuration affects the degree of impact. 
   The blue reference line shows the ideal performance with no defects.
   In both of the other cases, mode mismatch is introduced into by changing the curvature of the power recycling mirror by 2\,m ($\sim$0.1\,\%). 
   In red the recycling cavities are near-degenerate (round-trip Gouy phase $\sim$0\,deg), i.e. the higher-order modes generated by the mismatch are close to co-resonating with the target beam shape, and the transfer function is significantly altered.
   This implies that the sensitivity to gravitational-wave signals will be reduced.
   Meanwhile in black, the cavities are geometrically stable, so the alteration to the transfer function is negligible.\\
   Reproduced from \cite{Maggiore2024-thesis, Maggiore2025}. 
    }
\end{figure} 

We developed the \Finesse software\cite{Finesse1, finesse3-zenodo} to address the needs of the GW community. 
\Finesse is an open-source project, written in Python (with Cython), intended for both experts and non-expert coders to build interferometer models and interpret detector behaviours.
It enables calculation of optical phase changes of 1e-13, and interference effects of that level, for a multi-kilometer-scale optical system. 
It also enables us to account for the complex nature of GWDs by including optical, mechanical and electronic elements. 
While development is driven by GWD technologies, users can freely define their own optical systems. 
It thus finds usage both in detector characterisation and design, as well as broader applications.

\Finesse calculates the optical system as a nodal network, computing the steady-state optical couplings at each location. 
It is thus primarily suitable for quasi-static, linear optical systems, and for calculating transfer functions of those systems such as shown in Figure~\ref{fig:offsets}.
The optical field is described as a sum of higher-order Hermite-Gauss modes\cite{LivingRev}. 
Small/paraxial beam distortions can therefore be modelled as perturbations to the ideal Gaussian mode.

\section{An optically-focused approach to defining GWD mirror surface specifications}
As outlined in section~\ref{sec:Defects}, the requirements of GWD mirrors are defined by their impact on the ultimate performance of the detector. 
As we prepare for next-generation detectors CE and ET, these performance requirements are increasing, and new technologies are proposed.
Therefore we need to provide specifications that guarantee our performance requirements. 

Current GW optics are typically specified using a combination of statistical requirements, such as RMS and surface flatness \cite{Bingsley2017}. 
This may combined with measurements comparing to a targeted amplitude spectral density of the surface height (ASD), and/or Zernike coefficients. 
We have built new libraries to analyse mirror surface height data, referred to as 'mirror maps', calculating their ASD and Zernike polynomial content.
We compliment these well-established techniques\cite{phd.bond2014,Tao2020,LeJean2024} with an optical analysis of the mirror surface, using \Finesse to extract the higher-order optical mode content of the field reflected from the mirror surface, where the eigenmode of the relevant GWD arm cavity is the basis for the expansion. 
Finally, we have developed algorithms\cite{Bianchi2025} for generating large collections of `virtual' mirror maps which obey surface specifications based on ASD, Zernike polynomial, or combined requirements, with the aim of understanding how these metrics translate into optical performance for GWD design. 

\begin{figure} [tb]
   \begin{center}
   \begin{tabular}{c} 
   \includegraphics[height=5cm, trim={0 2cm 10.2cm 2cm},clip]{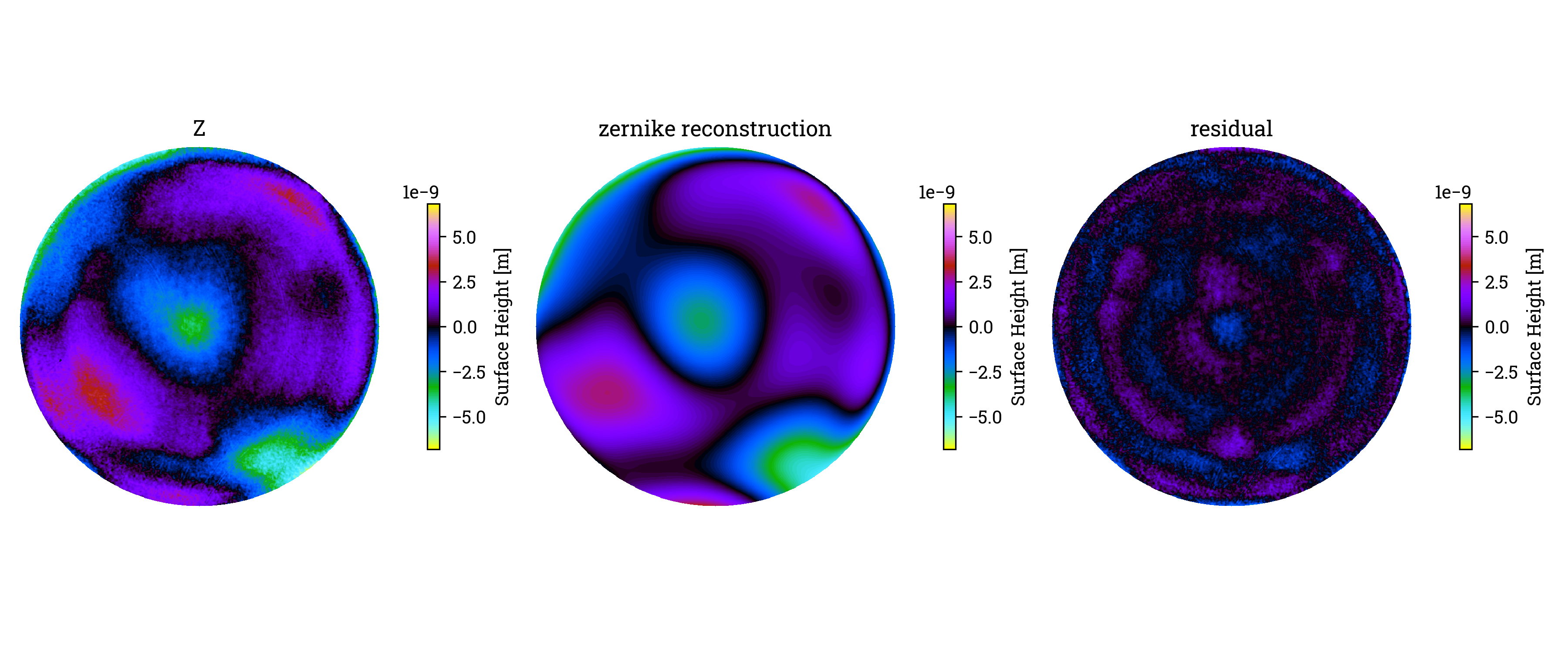 }
   \end{tabular}
   \end{center}
   \caption[Example Mirror] 
   { \label{fig:Mirror} 
   Left:
Map of surface height variation in a current GWD optic.
Light incident on this surface is imprinted with a non-uniform phase, which distorts the shape of the wavefront.
The optic has a diameter of 30\,cm.\\
    Right:
Reconstruction using Zernike Polynomials up to 8th order.
    }
\end{figure} 

\begin{figure} [tb]
   \begin{center}
   \begin{tabular}{c} 
   \includegraphics[height=7cm]{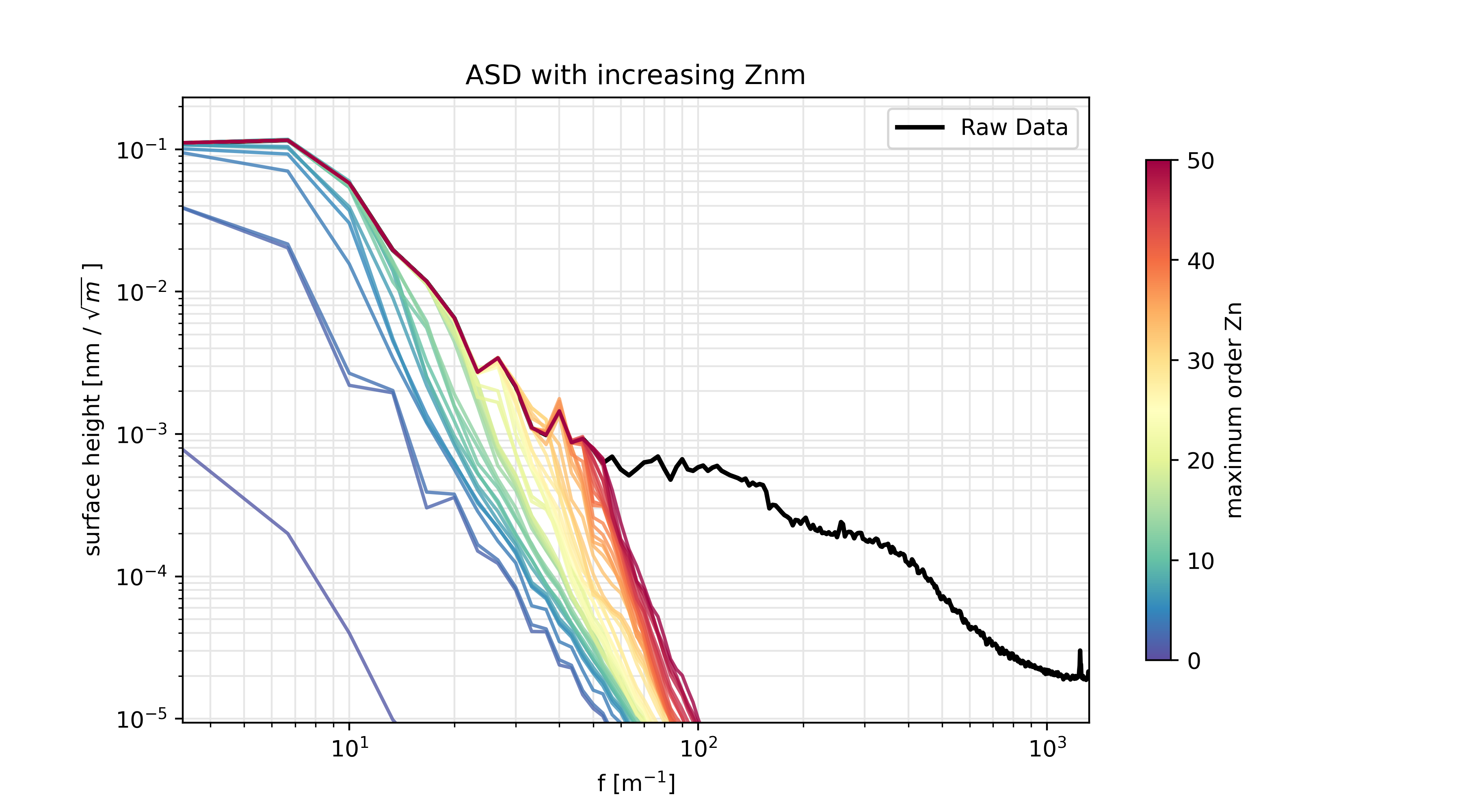}
   \end{tabular}
   \end{center}
    \caption[Established analysis techniques]
   { \label{fig:PSD-Znm}  
Amplitude spectral density of the mirror surface height (ASD), based on Fourier analysis, indicating the spatial frequencies composing the mirror surface. 
The black line shows the spectrum of the raw mirror data shown on the left of Figure~\ref{fig:Mirror}. 
The colourbar shows the ASD of the Zernike reconstruction of the surface for different maximum orders. 
Clearly, including more terms improves the quality of the fit, however there are diminishing returns as more orders are included.
    }
\end{figure} 

Figure~\ref{fig:Mirror} shows an example mirror map from a current GW detector. 
On the left is shown the original data, while the right depicts the reconstruction of the map, using Zernike polynomials up to 8th order.
The main features of the mirror have been captured, however the result is lacking in detail.
This is corroborated by Figure~\ref{fig:PSD-Znm}, which shows the corresponding ASD of the original surface, and of reconstructions of the surface up to 50th order.
As is typical in our field, the figure shows a one-dimensional ASD, averaging over the two-dimensional surface.
We see that the Zernike reconstruction matches the spectral density of the raw data well at low frequencies, so long as sufficient terms are included. 
However, the polynomial representation struggles to recover the high spatial frequency information, with diminishing returns are more orders are included in the reconstruction.

From these two figures, we can see the different information that is captured by the two methods: the polynomial representation retains 3-dimensional information about the surface, and, at low orders, can be intuitive to interpret as the terms correspond to well-known optical properties such as tilt, defocus, or astigmatism\cite{Zernike}. 
However, calculating large numbers of terms can be computationally demanding, and may still lack full detail of the optical surface.
Meanwhile, the ASD analysis loses some of the intuition for the overall shape of the mirror, but provides computationally efficient information about the smaller features.
Therefore we consider specifications that combine both approaches.

Preliminary results calculating the optical loss out of the fundamental Gaussian mode resulting from various virtual mirror maps\cite{Bianchi2025}, 
indicate that the different specification types change the constraints on the likely final loss. 
In particular, ASD-based maps have a wider range of loss values, 
while Zernike-based maps tend to have a significantly narrower distribution.
Additionally, once sufficient polynomial terms are included, the higher spatial frequency content appears to have limited impact. 

\begin{figure} [tb]
   \begin{center}
   \begin{tabular}{cc} 
   \includegraphics[height=5cm]{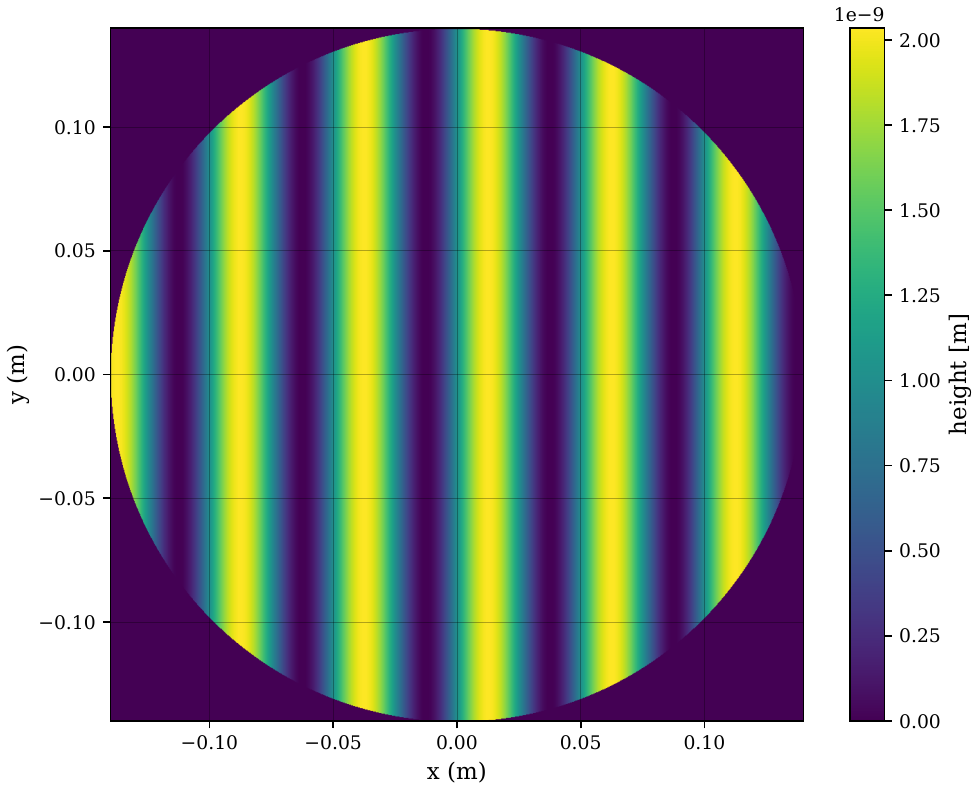} & 
   \includegraphics[height=5cm]{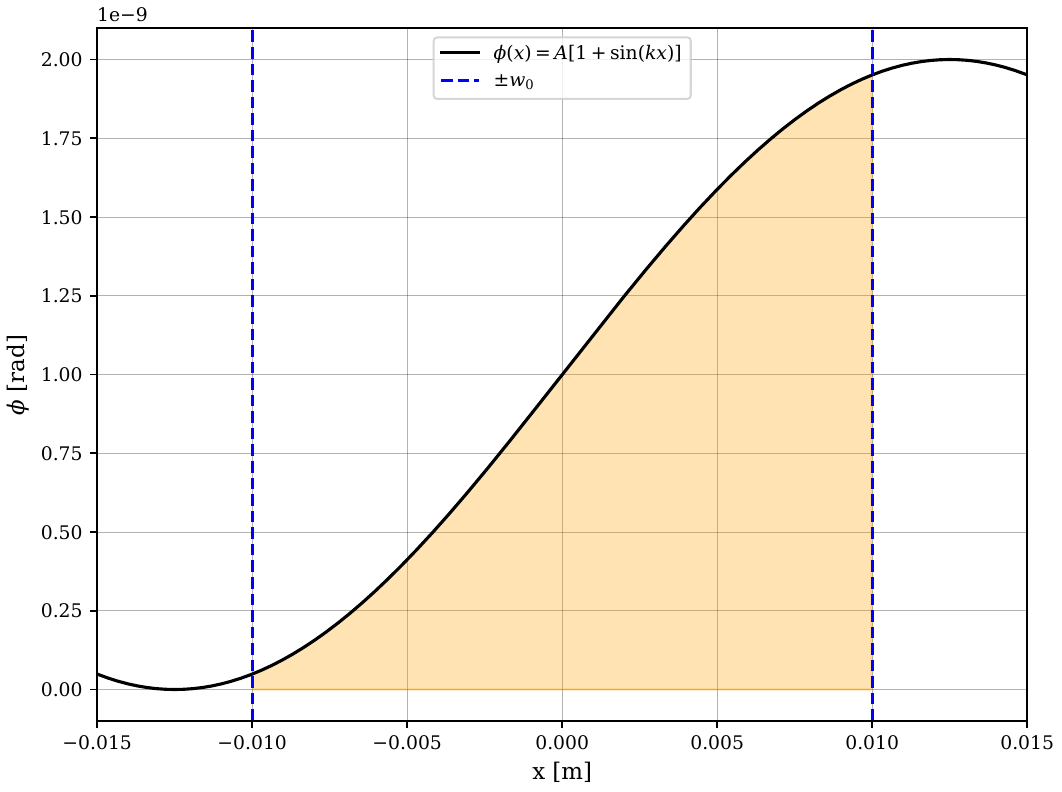} 
   \end{tabular}
   \end{center}
   \caption[Low frequency] 
   { \label{fig:FFT-spotsize-mirrorsize} 
   The local behaviour of low-spatial-frequency surface features dominates over their global description as a sinusoidal contribution to a spectrum.
   Here, an artificial mirror surface (left) appears to have a periodic form that would emulate a grating. 
   However, typically the beam size is significantly smaller than the optic in order to minimize clipping losses. 
   Therefore, the surface deformation experienced by the beam is primarily the local gradient (right), which for the example shown dominantly emulates a tilt rather than a diffractive grating.
    }
\end{figure}

Figure~\ref{fig:FFT-spotsize-mirrorsize} provides further insight into behaviour of ASD-based maps. 
 At low spatial frequencies the underlying assumption that only the amplitude of the defect is relevant is not longer valid.
The left figure depicts an artificial mirror surface containing a single spatial frequency\footnote{This represents a single point in an ASD plot; ASD analysis implies that a surface may be described as a superposition of such sinusoids.}, 15\,m$^{-1}$. 
While several periods of the sinusoid fit onto the mirror surface, the beam spot size in GWDs is typically chosen to be small enough to limit clipping losses to below 1\,ppm, thus the cavity eigenmode for this optic has a spotsize of just 0.02\,m - similar in dimension to the surface defect. 
Therefore, rather than experiencing a periodic pattern, the beam dominantly interacts with the surface shown on the right, i.e. it approximates a tilt. 
This highlights that at low spatial frequencies, the local gradient is much more critical to the optical performance than the global behaviour.
This is better constrained by the polynomial approach than the ASD method. 

We thus gain two threshold lengths for useful understanding regarding the interpretation of ASD plots such as Figure~\ref{fig:PSD-Znm} in the context of beam distortions: the beam spotsize, and the mirror diameter.

\begin{figure} [tb]
   \begin{center}
   \begin{tabular}{c} 
   \includegraphics[height=6cm, trim={0 0.3cm 18.5cm 1cm},clip]{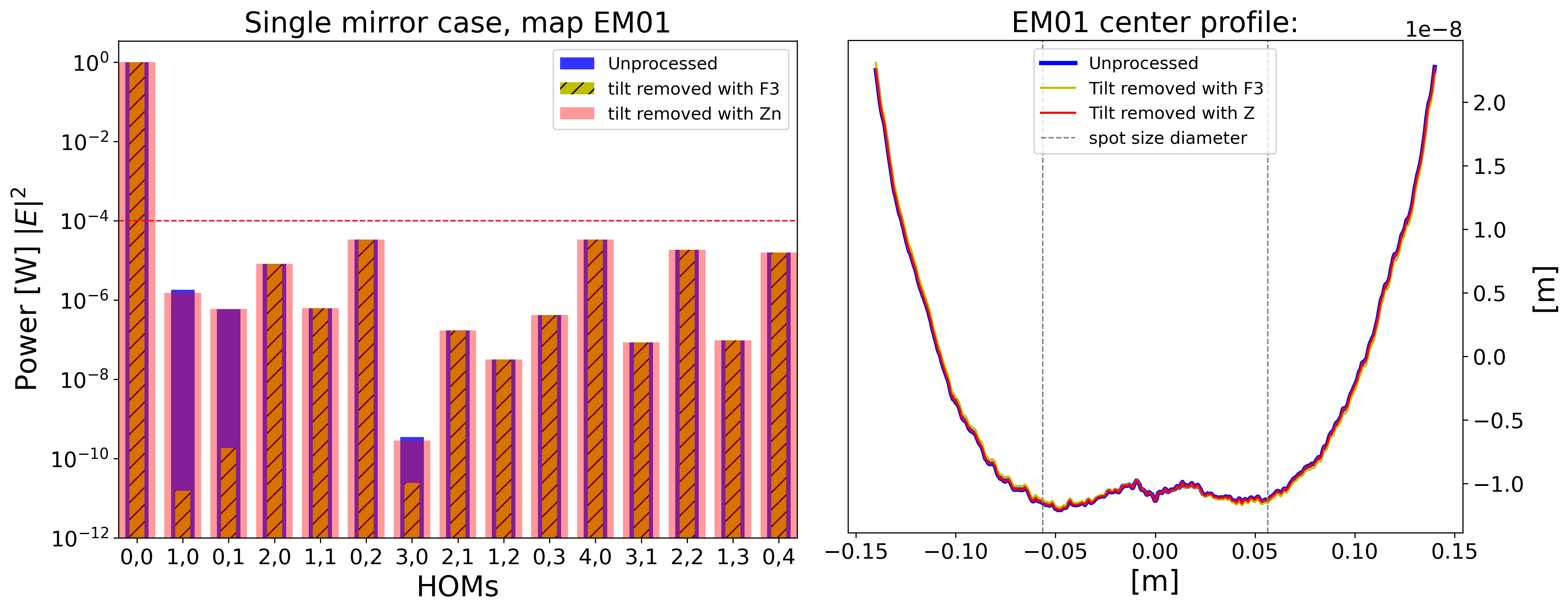} 
   \end{tabular}
   \end{center}
   \caption[Optical performance of map with different tilt-compensation methods] 
   { \label{fig:PTR} 
Distribution of optical power between the fundamental and higher-order optical modes, when different techniques are used to emulate the effect of optimal active alignment correction. 
Blue depicts the optical coupling to higher-order optical modes with no corrections. 
In pale red, the Zernike terms corresponding to tilt, $Z_{1,0}$ and $Z_{0,1}$, are directly removed from the map data.
Instead in yellow-striped, the linear terms, weighted by the size of the laser beam spot incident on the mirror, are removed.
This more closely matches the behaviour expected of the real control schemes used to correct these degrees of freedom, and results in a significantly 'cleaner' beam, regarding the higher-order optical mode content, than removal using Zernike polynomials.
Reproduced from \cite{Bianchi2025}.
    }
\end{figure}

Figure~\ref{fig:PTR} further provides insight into how to prepare maps for use in optical simulations.
Here, we plot the higher-order optical mode content, up to 4th order in Hermite-Gauss polynomials, for a GW mirror surface. 
Active controls in GWDs mean that, in practice, any small piston, tilt, or radius of curvature error that is present in the mirror surface will be corrected.
Therefore, it is appropriate to remove these contributions before assessing the mirror surface quality based on optical performance. 
Figure~\ref{fig:PTR} therefore shows the outcome of removing the tilt contribution to the surface, either by zeroing the corresponding Zernike polynomial coefficients, or by performing a \textit{weighted} removal of the linear surface component, based on the beam spot size. 
The latter method dramatically suppresses the first order higher-order optical modes, which dominantly couple to tilt. 
Therefore the spotsize-weighted approach is preferable for emulating the real optical system, and evaluating the spotsize-weighted contribution would give a more meaningful figure of merit for assessing the impact of the mirror surface tilt on the optical performance, compared to the Zernike coefficients.

\section{Conclusions and next steps}
Developing the final specifications for gravitational-wave detector optics that enable us to achieve our astrophysical and cosmological goals will depend on close collaboration between optics manufacturers and interferometer designers.

We have assessed ASD and Zernike-based techniques for qualifying optical surfaces for current and future gravitational-wave detectors, resulting in a clear understanding of their interpretation and limitations.
We confirm that the in-cavity optical performance is dominated by the low-spatial-frequency features of the surface, but find that while computationally efficient, spatial-frequency-based specifications do not sufficiently constrain the surface profile, as the local features of the surface dominate the optical performance. 
Polynomial-based specifications provide more control, however the Zernike polynomials are not always an optimal basis for describing mirrors in cavities with Gaussian beams. 
We are therefore now exploring alternative polynomials\cite{Feldmann2025}.

Beyond their current application to mirror surface defects, the tools and approaches we have developed for this work can now be applied to the broader suite of optical defects, including substrate inhomogeneities and dynamic (e.g. thermal) behaviours.



\acknowledgments 
We thank J. Degallaix for useful discussions. 
This publication is part of the project “Smoothing the Optical Bumps in the Road for Future Gravitational-Wave Detectors”, led by A.~C.~Green, with project number VI.Veni.212.047, which is financed by the Dutch Research Council (NWO).

\bibliography{main}
\bibliographystyle{spiebib} 

\end{document}